\documentclass[acus]{JAC2000}

\usepackage{graphicx}

\begin{document}
\title{NEW METHOD OF DISPERSION CORRECTION IN THE PEP-II LOW ENERGY 
       RING\thanks{Work supported by the Department of Energy under 
       Contract No. DE-AC03-76SF00515}}

\author{I. Reichel, Y. Cai, SLAC, Stanford, California\\
        SLAC-PUB- physics/0008216}

\maketitle

\begin{abstract} 
The sextupole magnets in the Low Energy Ring (LER) of PEP-II are
grouped in pairs with a phase advance of 180~degrees. Displacing the magnets
or moving the orbit to displace the beam in the magnets in
an antisymmetric way creates a dispersion wave around the ring. This
can be used to correct the vertical dispersion in LER without
changing the local coupling. Results from
simulations are shown.
\end{abstract}

\section{INTRODUCTION}

The luminosity of PEP-II is currently mainly determined by the vertical 
beam size at the interaction point (IP). In the Low Energy Ring (LER)
the vertical beam size at the IP is, to some extent, caused by the 
residual vertical dispersion in the ring. It is hoped that by lowering the
dispersion in the ring, the vertical beam size at the IP can be decreased 
and the luminosity therefore increased.

The sextupoles to correct the chromaticity in LER are grouped in 
non-interleaved pairs of the same strength. Therefore moving one of 
the sextupoles of a pair up (or moving the beam in the sextupole 
using a closed 
orbit bump) and the other one of the pair down creates a dispersion 
wave around the ring without affecting the coupling or the orbit 
outside the region. We want to use one or more of these dispersion 
waves to try to cancel some of the residual vertical dispersion in
the ring in order to minimize the vertical beam size at the IP.

\section{SIMULATION}

\subsection{LEGO}

The simulations are done using LEGO~\cite{LEGO}. Five different seeds 
are used for the misalignment. All five seeds give RMS dispersions and 
orbits of the size that is observed in the real machine. Orbit and 
dispersion are corrected using the same algorithms that are used in the 
control room. The coupling is minimized using the closest tune approach
in a way similar to the procedure used in the control room on the 
real ring.

\subsection{Effects of Sextupole Alignment}

\begin{figure}[htb]
\centering
\includegraphics*[width=60mm]{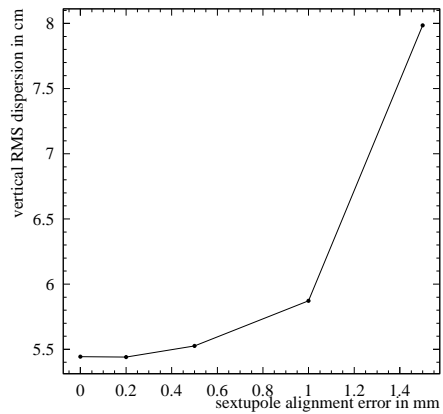}
\caption{Dependence of the residual vertical dispersion on the 
vertical sextupole alignment (average over five seeds).}
\label{F:disp}
\end{figure}

For a vertical alignment error of $0.5\,$mm for the
sextupoles (this is assumed to be the 
error in the machine), the residual vertical dispersions for these seeds 
are between 5 and $6\,$cm. Using an error of $1\,$mm, the dispersions 
are only slightly larger for four seeds and grow from 6 to almost 
$7.2\,$cm for one seed. Figure~\ref{F:disp} shows the dependence of 
the dispersion on the average error.
average.

\subsection{Moving Sextupole Pairs}

In the simulations we currently move the sextupoles by changing their 
alignment. This is easier and "cleaner" than a closed orbit bump.

A sextupole pair is chosen. The simulation program then loops over 
position changes from $-5$ to $+5\,$mm in steps of one~mm. One of the 
two magnets is moved up by the appropriate amount, the other one down 
(taking into account the original (mis-)alignment of the magnet).

At each step the dispersion and the vertical beam size at the IP
are calculated. The vertical beam size is calculated using the 
algorithm described in~\cite{ybeamsize}.

The correction algorithm was studied for vertical
sextupole misalignments of $0.5$ and 
$1.0\,$mm, where the average dispersion for the five seeds is
$5.5$ and $5.9\,$cm respectively. The average beam sizes
at the IP are $3.52$ and $4.11\,\mu$m.
For each calculation two different sextupole pairs and the same 
five different seeds as before are used.

\section{RESULTS}

\begin{figure}[tb]
\centering
\includegraphics*[width=60mm]{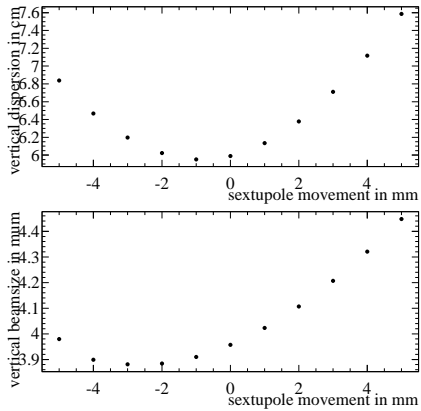}
\caption{Simulation result for one sextupole pair and one seed.}
\label{F:scan}
\end{figure}

Figure~\ref{F:scan} shows the results of a typical simulation 
run. The dispersion and vertical beam size at the IP are plotted 
versus the alignment change of the sextupole pair. One can see 
that the two parameters have their minimum not at the same 
misalignment.

From each the dispersion and the beam size one can obtain an optimal 
position for the respective sextupole pair (see Fig.~\ref{F:scan}). 
However, the results of 
the two don't always agree with each other. 

Moving one sextupole pair such that the minimum dispersion is 
obtained, the vertical beam size on average shrinks slightly to $3.51$ and 
$4.07\,\mu$m respectively (to be compared to $3.52$ and $4.11\,\mu$m).
In this case the sextupoles have to be moved on average by $0.8$
and $1.2\,$mm respectively.

Using the minimum of the vertical beam size one can obtain slightly 
better results: $3.50$ and $4.06\,\mu$m respectively.

\section{CONCLUSION}

Unfortunately for this method, the general steering algorithm in PEP-II is
good enough to obtain small dispersions and small vertical beam sizes. 
Moving sextupole pairs can decrease the vertical beam size only by
about $0.1\,\mu$m, which is of the order of two to three percent.
Using this method on the real machine is made complicated by two things:
In the machine closed orbit bumps have to be used instead of moving the 
magnets themselves and it is not very easy to optimize on the vertical 
beam size at the IP itself which seems to be the method to be used 
giving the difference in location of the minima. 

Nevertheless the method might be useful on the real machine as the
dispersion correction in the steering package currently works not
very well. This method might be faster and more efficient. This will
be studied.


\begin{thebibliography}{9}

\bibitem{LEGO}
Y. Cai et al.: LEGO: A Modular Accelerator Design Code, Proceedings of the 
17th IEEE Particle Accelerator Conference, Vancouver, Canada, 1997.

\bibitem{ybeamsize}
Y. Cai: Simulation of Synchrotron Radiation in an Electron Storage Ring, 
Proceedings of the 15th Advanced ICFA Beam Dynamics Workshop 
on Quantum Aspects in Beam Physics, 1998. SLAC-PUB-7793
 
\end{thebibliography}
\end{document}